# EXPERIMENTAL INVESTIGATIONS OF UNDERWATER AND AIRBORNE NOISES PRODUCED BY A LARGE HOVERCRAFT IN URAL RIVER ESTUARY

A.I. Vedenev[1,a]; O.Yu. Kochetov[1,b]; A.A. Lunkov[2,c]; A.S. Shurup[1, 3, 4,d] ; S.S. Kassymbekova[5,e]

[1] *Shirshov Institute of Oceanology of the Russian Academy of Sciences. 36 Nakhimovsky Ave., 117997 Moscow, Russia*

[2] *Prokhorov General Physics Institute of the Russian Academy of Sciences. 38 Vavilov Str., 119991 Moscow, Russia*

[3] *Schmidt Institute of Physics of the Earth of the Russian Academy of Sciences, B. Gruzinskaya Str., 123242 Moscow, Russia*

[4] *Lomonosov Moscow State University, Faculty of Physics, Leninskie Gory, Moscow 119991, Russia*

[5] *KMG Systems&Services LLP, 26-84 Koshkarbayev Str., 010000 Astana , Kazakhstan*

E-mail: [a]*vedenev@ocean.ru,* [b]*realspinner@gmail.com,* [c]*lunkov@kapella.gpi.ru,* [d]*shurup@physics.msu.ru,* [e] kassymbekova@kmgss.kz

Simultaneous measurements of underwater and airborne noises produced by Griffon BHT130 hovercraft were carried out in environmentally sensitive area – wildlife preserve in the area of the Ural River estuary near the Caspian Sea shelf. Measurements were organized to assess the possible negative impact of noise from hovercraft on fish and birds in wildlife preserve. The particle velocity of underwater noise was estimated by using a gradient-type vector receiver. That was a distinctive aspect of the underwater noise studies since the majority of fish perceives the sound in terms of vibration of particles, and only a few as the pressure. Using synchronous recording of underwater and airborne noises, the mutual correlation of these data was investigated. The obtained correlation levels between underwater and airborne noises produced by hovercraft can be used for simplified estimation of the upper boundary of underwater noise level by measuring levels of airborne noise. The measured and estimated maximal levels of underwater noises of hovercraft are considerably lower than noises from conventional vessels with underwater engines, that makes hovercraft attractive alternative for use in locations with high underwater noise requirements, such as Ural River estuary and Caspian Sea shelf.

*Keywords:* hovercraft, underwater noise, airborne noise, particle velocity.

PACS: 43.50.Lj, 43.50.Rq, 43.30.Nb

# 1. Introduction

Underwater anthropogenic noise has increased significantly over the past few decades due to the continuous growth in the number of ships used for logistical tasks [*Smith T. A., Rigby J.* 2022]. There are forecasts that underwater anthropogenic noise levels will double over the next 11.5 years [*Jalkanen, J. P., Johansson, 2022*], and the growing number of research emphasizes the negative impact on aquatic ecosystems that such growth is already having today [*Smith T. A., Rigby J.* 2022; *Jalkanen, J. P., Johansson, 2022*]. Any use of ships in environmentally sensitive areas, where it may result in negative impacts on the existing ecosystem, requires comprehensive noise studies. Special attention is paid to this problem in the development and operation of offshore gas and oil fields [*McKenna et al., 2012; Jimenez-Arranz et al., 2008; Rutenko A. N., Ushchipovskii V. G., 205*], with the main attention to the study of noise of ships with traditional propeller engine located underwater, as well as the negative impact from pile driving [*MacGillivray and Racca, 2006; McCauley et al., 2021*]. At the same time, to solve logistic tasks, ships of a new type – hovercrafts – are more and more frequently used. The movement of hovercrafts is determined by two types of engines: lift engines, providing elevation above the water surface, and propulsion engines, forming thrust in the horizontal plane. Since a hovercraft does not actually touch the surface of the water layer when moving, the underwater noise generated during such movement is noticeably less than the noise from vessels with underwater propulsion system [*Blackwell S. B., Greene Jr C. R., 2005*]. This makes hovercraft attractive for use in areas with special requirements for environmental regulations on underwater noise. [*Roof C.J., Fleming G.G., 2001*]. It should be noted that airborne noise generated by a moving hovercraft can be significant, which requires simultaneous measurement of airborne and underwater noise in studies of anthropogenic impact on ornio- and ichthyofauna [*Blackwell S. B., Greene Jr C. R., 2005; Vedenev A.I., Lunkov A.A. 2018, 2023*]. Noise studies of hovercrafts are currently being emphasized due to the possibility of their application both on the surface of water and on the surface of land or ice, which makes this type of vessel largely universal in terms of regions of their possible application.

This paper reports the results of in-situ measurements of airborne and underwater noise of the Griffon Hoverwork BHT130 hovercraft, named "*Caspian Falcon* ". Measurements were carried out in the Estuary of the Ural River and the North Caspian of the Kazakhstani sector (Fig.1), which are a unique wildlife preservation area, nesting area for rare birds, as well as a spawning place for valuable food fish can be attributed as one of such regions. On the other hand, the largest offshore oil field Kashagan is located in this region, where the oil companies have started using the hovercrafts on a regular basis for the transportation of personnel and cargo to a marine platform in the North Caspian

Sea along the routes in the estuary of the Ural River. Conducting research aimed at measuring underwater and airborne noise levels in this region is an urgent and highly demanded task.

Earlier, in [*Vedenev, Lunkov et al. 2023*] it was shown that underwater noise from *Caspian Falcon* hovercraft turns out to be very small in comparison with vessels equipped with underwater engines. At the same time, air noise levels can reach significant values, while remaining within acceptable levels in terms of negative impact on ornithofauna. The main difference of the present work from the previously presented results [*Vedenev, Lunkov et al. 2023*] is the joint analysis of air and underwater noise. On the one hand, it has advanced understanding of the mechanisms of underwater noise generation form hovercraft. Thus, it turned out that at approaching of the hovercraft to the observation point the main contribution to the underwater noise is given by the lift engine noise. Whereas the impact of propulsion engine noise is predominant when hovercraft moving away from the observation point. On the other hand, the obtained values of air and underwater noise correlation allow us to propose a simplified scheme for estimating the maximum levels of underwater noise on the basis of air noise measurements (such estimates are valid for a given ship and research region, i.e. for the considered source and conditions of acoustic signal propagation).

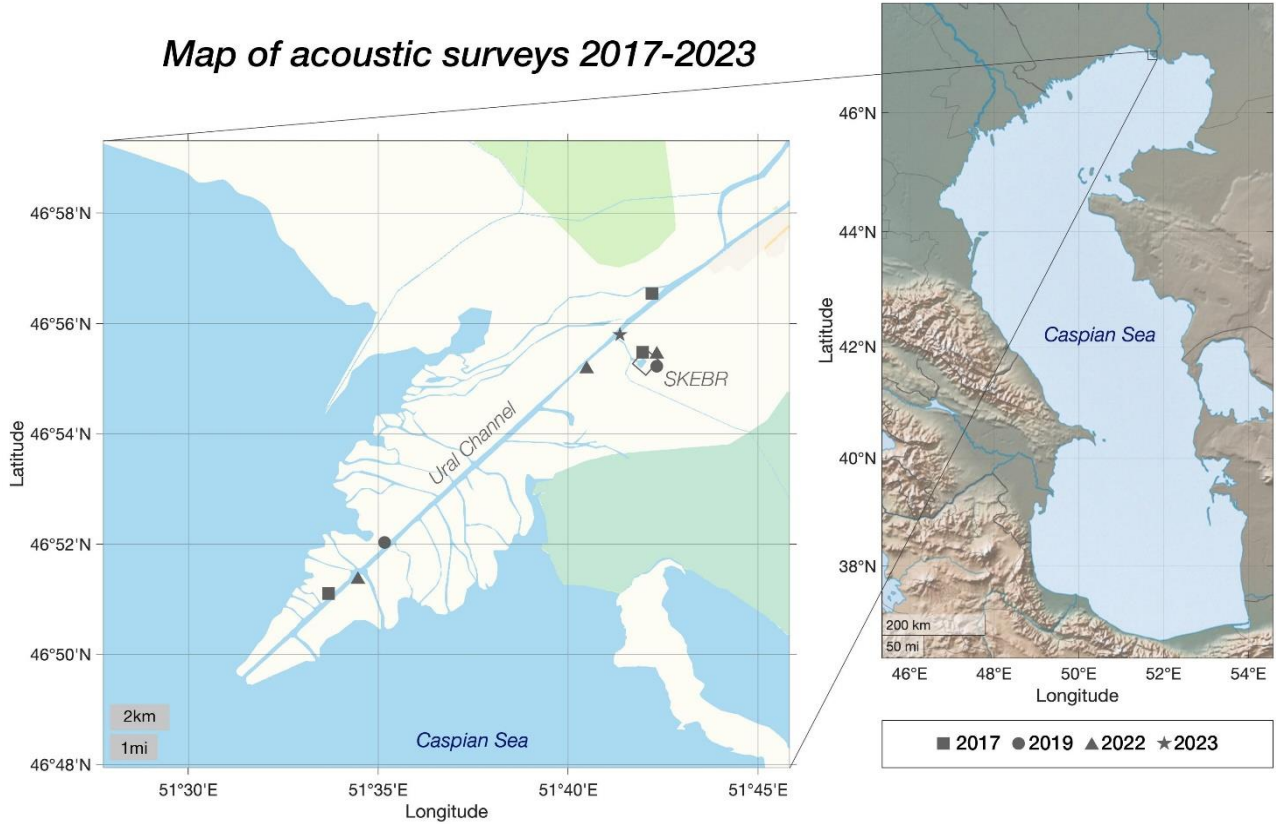

Fig. 1. Map of field measurements held in 2017, 2019 and 2022 in the North Caspian region (a) focused on *Caspian Falcon* airborne and underwater noise investigation.

The fact is that underwater measurements are significantly more expensive and complex than airborne noise measurements, so simplifying the estimates of underwater noise levels is an important task. The results obtained in this work can be a basis for the new scheme of hovercrafts noise investigations with simplified requirements for its practical realization.

## 2. Materials and methods

Specifications of the investigated hovercraft *Caspian Falcon* (Fig. 2) are as follows:

- Type: Griffon Hoverwork BHT130 hovercraft;
- Dimensions: Length: 30 m; beam: 13 m;
- Power system: Four diesel engines: two lift engines (900 HP each); two propulsion engines (1200 HP each);
- Propellers: two 6-blade propellers.

This is quite large hovercraft, in comparison with a small-sized Griffon 2000TD hovercraft, studied in [*Blackwell and Greene, 2005*], a Griffon BHT130 hovercraft investigated in this work is three times larger.

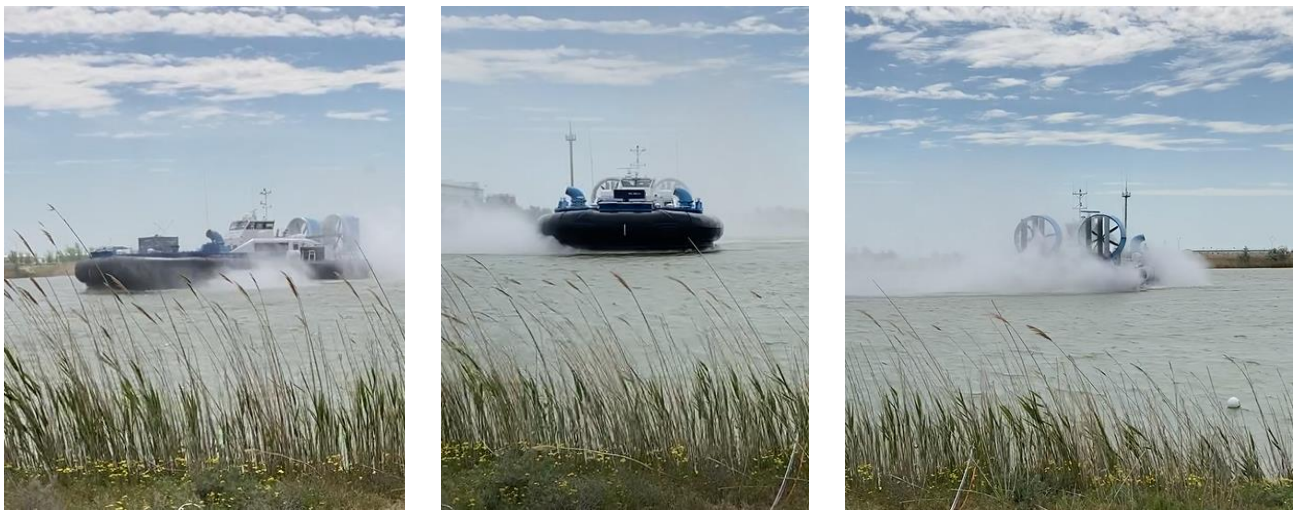

Fig. 2. Different views of the hovercraft *Caspian Falcon* during measurements.

Field experiments were carried out in the Ural-Caspian Channel (a man-made riverbed of the Ural River) in July-August 2017, August 2019 and May 2022 [*Vedenev, Lunkov et al. 2018; Vedenev, Kochetov et al. 2023; Research Report, KMG Systems & Services 2019; Summary Report, KMG Systems & Services 2022*] at various sites: the NCOSRB enclosed basin (the North Caspian Environmental Oil Spill Response Base), the Ural-Caspian Channel and on the North Caspian shelf of Kazakhstan (Fig. 1). The Ural-Caspian Channel has almost abrupt shores and an average depth of ≈ 5 m and width of ≈ 140 m. Water depth of the Caspian Sea near the Ural River estuary is ≈ 2 m.

A hovercraft airborne noise has been measured in accordance with an ISO standard [*ISO 2922:2020*]. However, very shallow water (< 5 m depth) conditions of the Ural-Caspian Channel and the North Caspian shelf do not fit the requirements of the ISO standard for underwater noise measurements. ISO 17208-1:2016 requires the water depth of 150 m [*ISO 17208-1:2016*]. Therefore, the measured underwater noise parameters of the hovercraft relate to this specific region only.

To measure underwater noise, HTI-96min hydrophones were used, as well as a hydrophone manufactured at the Shirshov Institute of Oceanology (SIO hydrophone). Particle velocity measurements of hovercraft underwater noise were conducted using a pressure gradient sensor (PGS) which consists of four non-coplanar HTI-96min [*Vedenev, Kochetov et al. 2023; Vedenev, Lunkov 2023 et al. 2023*]. Finite-difference approximation of the spatial derivatives of the pressure field measured by these hydrophones makes it possible to assess three orthogonal projections of the particle velocity [*Crocker M. J., Arenas J. P. 2003; Nedelec S. L., Ainslie M. A. et al. 2021*]. In [*MacGillivray A., Racca R. 2006*], a similar PGS was used to study the impulse noise produced by the pile driving.

Airborne noise was measured using a calibrated station NTM-Zashchita noise meter with an MK-265 microphone and a windscreen used to suppress flow-noise caused by wind. In this case, the values of acoustic noise levels in different frequency bands were measured. Additionally, measurements were taken with a video camera, the sound from which made it possible to obtain acoustic pressure values as a function of time. These data were used to construct cross-correlations between underwater and airborne noise. Before calculating correlations, airborne and underwater noise recordings were synchronized with the required accuracy. Airborne noises were measured by microphones mounted on a tripod on the ground 2 to 4 m above the water surface. In experiments the average wind speed was additionally controlled to be less than 6 m/s.

During the measurements the hovercraft was running at different speeds along buoy lines (Fig. 3). Measuring equipment was installed near this line at the bottom of the Ural River or on the shore. The use of buoy lines allowed the hovercraft captain to set a trajectory of movement near the measuring equipment that was close to a straight line. At the same time, the speed of the hovercraft was kept as close to a constant as possible. A GPS tracker and a laser rangefinder were used to monitor the hovercraft position relative to the recording equipment. Wind speed was measured with an anemometer. The more or less similar setup was used for all measurements over different years. At the same time, the equipment used changed. For example, particle velocity measurements using PGS were not carried out until 2022.

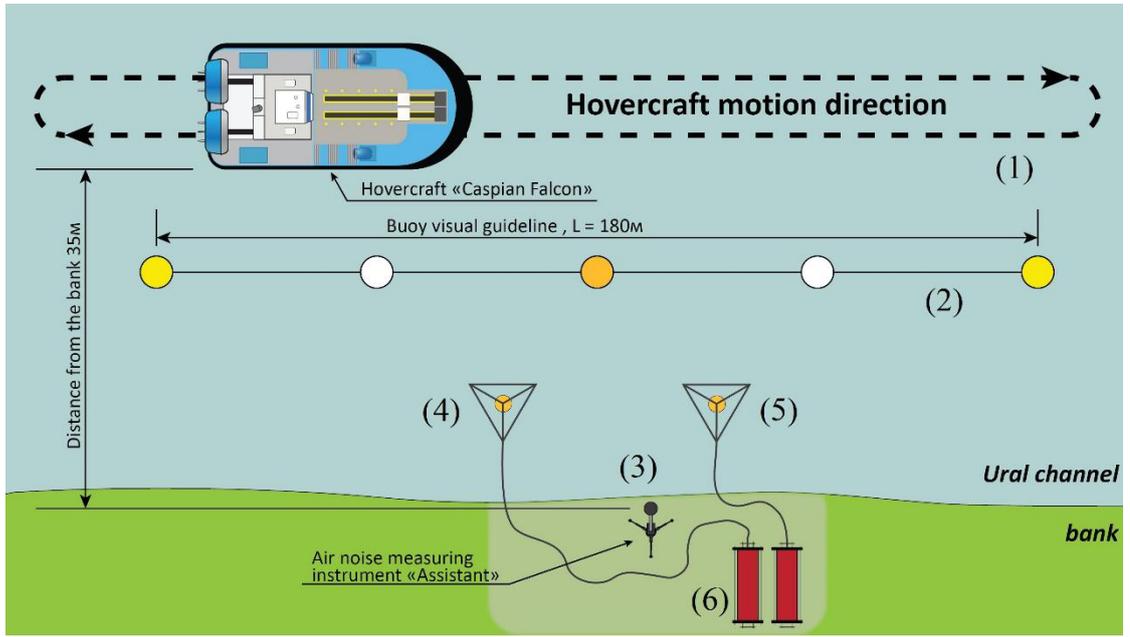

Fig. 3. Example of experimental setup of the noise measurements in the Ural-Caspian Channel: hovercraft tracks (1) were controlled by buoy lines (2); airborne noise was measured on the shore (3); underwater noise was measured by 2 independent devices (4), (5) to increase the probability of success of underwater measurements and to obtain independent measurements for self-checking. Different devices were used in different years, for example in 2019 single HTI-96min and SIO hydrophones were used and in 2022 PGS and Bruel&Kjaer 8104 hydrophone with autonomous recording system were applied. Underwater noise recording systems were located on shore (6).

2.1. Metrics of underwater and airborne noises.

The following metrics are calculated from pressure time series $p_j(t)$ recorded at each hydrophone $j$:

- Time dependence of the sound pressure level (SPL)

$$SPL(t) = 20 \lg \left( \frac{\sqrt{\frac{1}{\Delta t} \int_t^{t+\Delta t} <p^2(t')> dt'}}{p_0} \right), \quad (1)$$

measured in dB re $p_0 = 1 \mu Pa$, $\Delta t = 1$ s is the averaging time, $<..>$ means an averaging over four hydrophones of PGS, i.e., $<p^2(t')> = \frac{1}{4}\sum_{j=1}^{4} p_j^2(t')$ (if only one hydrophone was used there were no averaging);

- Time dependence of the particle velocity level (VL)

$$VL(t) = 20 \lg \left( \frac{\sqrt{\frac{1}{\Delta t} \int_t^{t+\Delta t} v^2(t')dt'}}{v_0} \right) \quad (3)$$

measured in the units of dB re $v_0 = 1$ nm/s. Here, $v(t') = \sqrt{v_x^2(t') + v_y^2(t') + v_z^2(t')}$ is the absolute value of particle velocity at time $t'$. Components $(v_x, v_y, v_z)$ of the particle velocity vector are calculated by taking the difference of the sound pressure at the relevant pairs of hydrophones of PGS. For example, the projection of particle velocity $v_{12}(t)$ onto the line 1-2, connecting two hydrophones (#1 and #2) is defined by the formula:

$$v_{12}(t) = -\frac{1}{\rho h} \int_0^t (p_1(t') - p_2(t'))dt', \quad (4)$$

where $\rho$ is the density of water; $h$ is the spacing between hydrophones #1 and #2. In the frequency domain, Equation (4) takes the form [*Nedelec S. L., Ainslie M. A. et al, 2021*]:

$$v_{12}(f) = \frac{i}{2\pi f \rho} \frac{p_1(f) - p_2(f)}{h} \quad (5)$$

where $i$ is the imaginary unit. Further, the maximal levels of sound pressure level and particle velocity level will be analyzed.

For airborne noise the sound pressure level (SPL) were calculated followed the recommendations of the ISO standard [*ISO, Airborne sound, 2021*]. Time series of the 1 s-averaged A-weighted broadband (20-20000 Hz) levels in dBA, and Z-weighted broadband (20-20000 Hz) levels in dBZ, were recorded at every passage of the hovercraft. The similar to underwater metrics, the maximal values of airborne sound pressure levels will be analyzed.

To compare the measured maximal levels of the hovercraft airborne and underwater noises with the literature data, other vessels or between different passages of the hovercraft in different years, the maximum levels are normalized to a reference distance of 30 m. To calculate the normalized level $SPL_{30}$, the spherical decay law is used:

$$SPL_{30} = SPL(t_{cpa}) + 20 \lg \left( \frac{r_{cpa}}{30} \right), \quad (6)$$

where $t_{cpa}$ and $r_{cpa}$ are the values of time and range at the point of closest approach (CPA). The spherical spreading of an acoustic wave recorded by microphone in air and by a close to bottom mounted hydrophone was verified experimentally.

### 3. Results
3.1. Underwater noise.

Fig. 4 shows results of measurements of the maximum pressure levels (SPL) and particle velocity (VL) for the underwater noise of the hovercraft, measured at different speeds of its movement. As it can be seen in Fig. 4 at low speed (up to 7 m/s) and speeds close to maximum values (more than 15 m/s) the underwater noise levels exceed noise levels at cruising speeds (7-15 m/s). This effect is explained by the fact that at low speeds (low rotations of lifting and propulsion engines) the hovercraft is located near the surface of the water, and may even touch its surface. In this case, the vibrations of the hovercraft hull, as well as the noise of the lifting engines, provide a high level of underwater noise. At the maximum speeds of the hovercraft motion (i.e. at the highest rotation speeds of its engines), airborne noise increases greatly, which contributes to underwater noise. At average speeds, or cruising speeds, an intermediate situation is observed: the noise of the hovercraft engines is not maximum, while the hovercraft moves without practically touching the surface of the water. Thus, the average speeds of hovercraft movement (but not the minimum, as one might assume) are the most preferable in terms of hovercraft impact on the ichthyo- and ornithofauna of the Ural river.

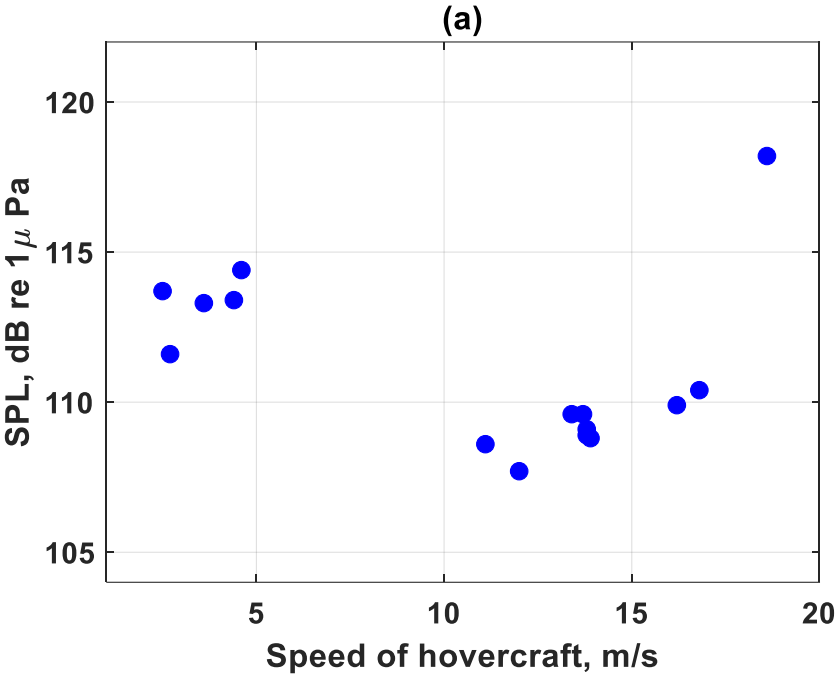

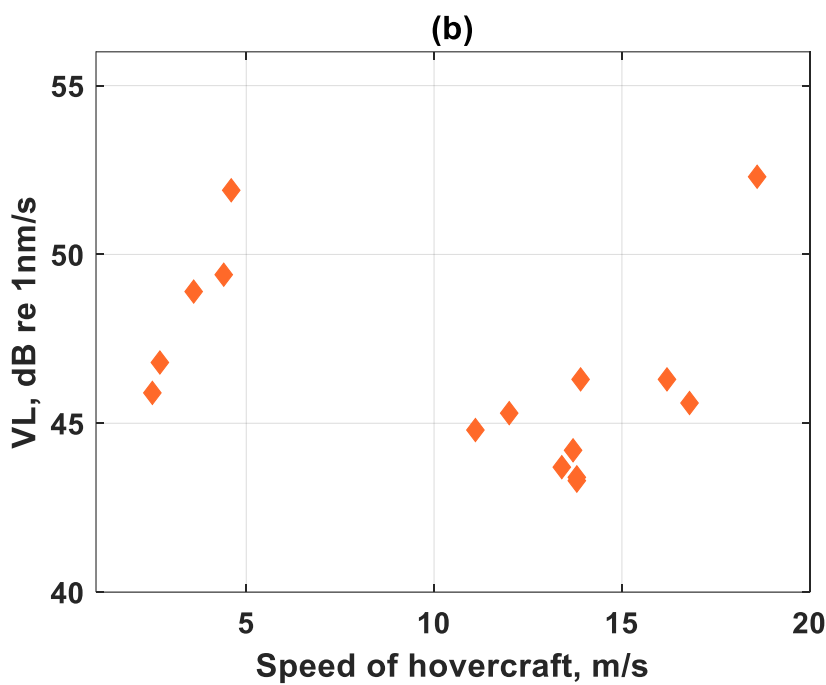

Fig. 4. Dependences of maximum pressure levels (SPL) and particle velocity (VL) of underwater noise measured when hovercraft Caspian Falcon was moving at different speeds. Levels are given for a distance of 30 meters.

3.2. Correlation between airborne and underwater noise

Since a hovercraft has no underwater parts, underwater noise can be produced either by the airborne noise from propulsion engine transmitted through the air-water interface (1st mechanism) or by noise from lift engines which forms strong air flows beneath the hovercraft and can cause cavitating bubbles generated beneath its skirt (2nd mechanism). The role of both mechanisms in forming the total noise field should be estimated.

Fig. 5 shows spectrograms of airborne and underwater noise for the hovercraft passing at the maximum speed. Vertical white dashed lines denote time at the CPA. The spectrum of airborne noise consists of persistent narrow spectral peaks (tones) during both approach and retreat (see also blue lines in Fig. 6). Narrow peaks correspond to the fundamental frequency and its multiples associated with thrust propeller. However, tone frequencies change due to Doppler shift. In underwater noise, spectral peaks are present only during the retreat (see Fig. 5(b-c) and Fig.6). When approaching the CPA, underwater noise spectrum looks rather smooth (no distinct peaks). Probably, it means that during the approach, the bubble cavitation noise (2nd mechanism) dominates. The noise from the thrust propellers is shaded by the hovercraft cabin. During the retreat, on the contrary, there is no shading, and the effect of airborne noise transmission into water is more pronounced.

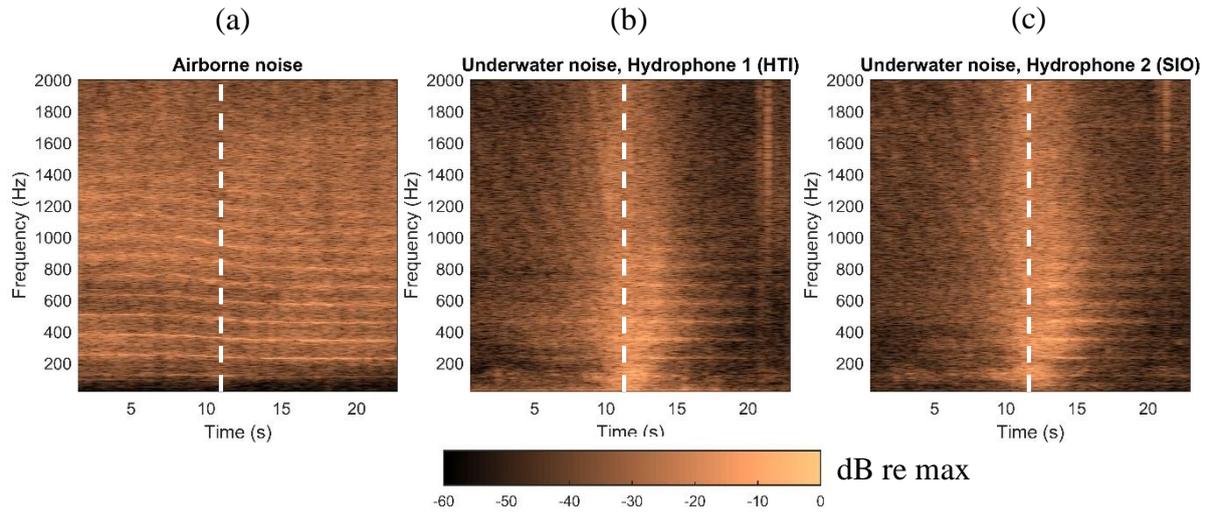

Fig. 5. Normalized spectrograms of (a) airborne and (b-c) underwater noise at maximum speed. White dashed lines denote the CPA.

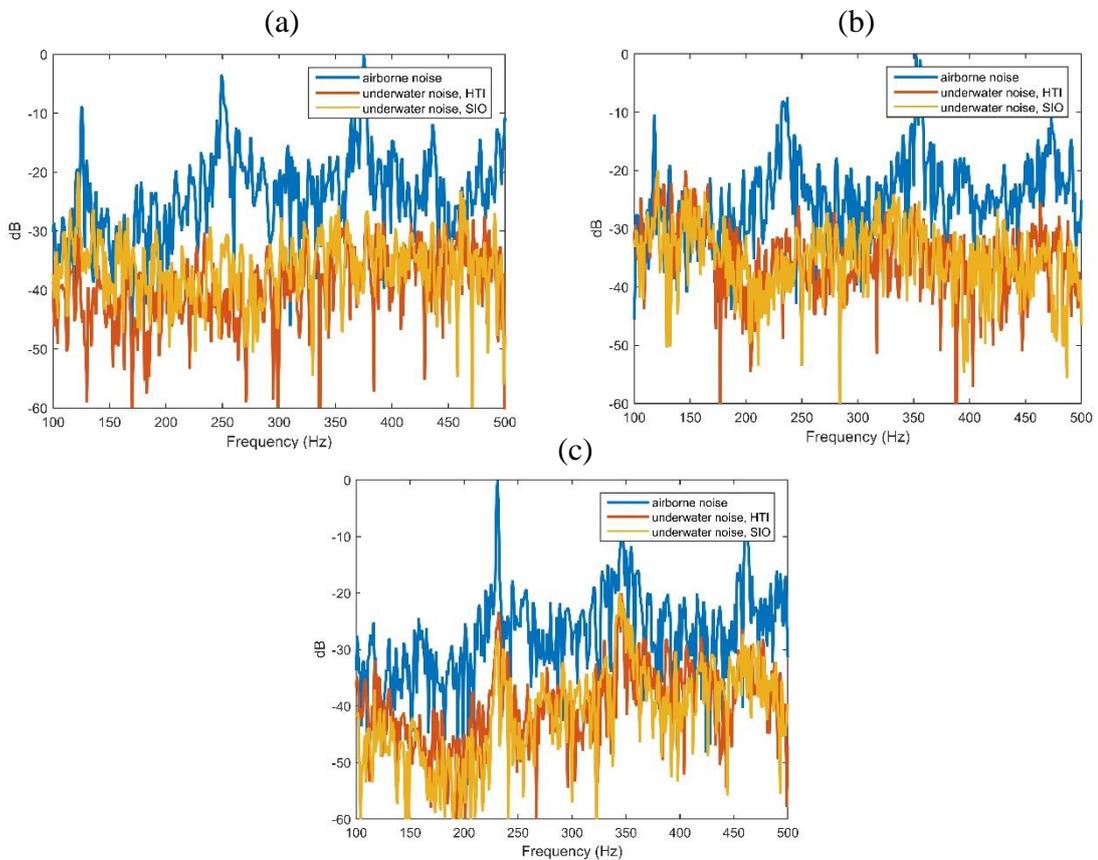

Fig. 6. Normalized noise spectra comparison (a) before passing the CPA (approach), (b) at the CPA, (c) after passing the CPA (retreat).

Some verification of this hypothesis can be obtained by analyzing the correlation between underwater and airborne noise. For this purpose, noise segments of 1 s duration are correlated. Time shift to the next segment is 0.5 s. Frequency band is 100 to 2000 Hz. The time evolution of the correlation function envelope is demonstrated in Fig. 7. One can see that maximum correlation corresponds to the retreat of the hovercraft, and no correlation is found for the approach.

Another option of correlation analysis is to compare maximum sound pressure levels of airborne and underwater noise at different speeds at a nominal range (30 m). Fig. 8 shows underwater noise level plotted versus airborne noise level in dBA or dBZ. Each point corresponds to one passage of a hovercraft. The statistics covers all the regimes: low speed, cruising speed, and maximum speed. Linear regression model (yellow dashed line in Fig. 8) fits well the observed relation between noise levels, with the coefficient of determination $R^2 = 0.64$-$0.68$. It means that the maximum level of the underwater noise can be predicted from the maximum level of the airborne noise. It also implies that the maximum level of underwater noise is produced by the 2$^{nd}$ mechanism.

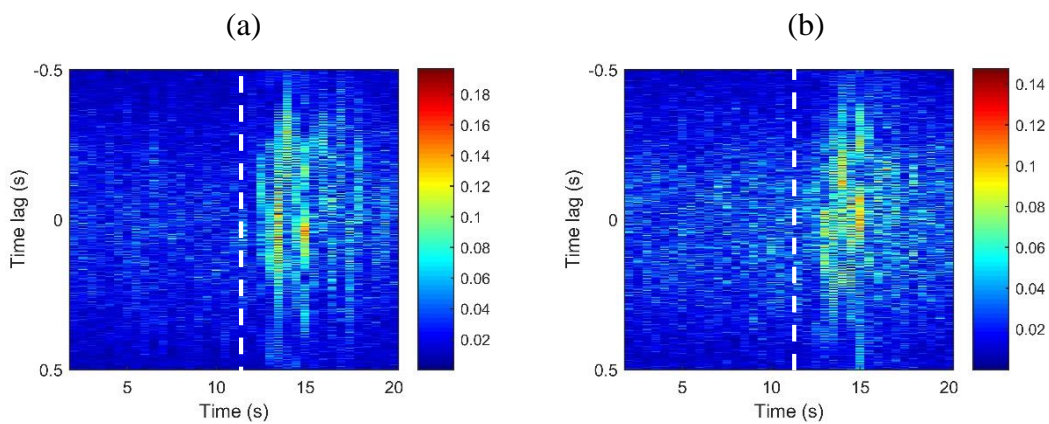

Fig. 7. Correlation function between airborne and underwater noise. (a) – hydrophone HTI, (b) – hydrophone SIO. White dashed lines denote the CPA.

## 4. Discussion and conclusions

The measured correlation of airborne and underwater noises allows to propose a methodology of simplified estimation of the upper boundary of underwater noise level by measured values of airborne noise of hovercrafts. Taking into account the complexity of field work to determine the impact of acoustic noise of hovercrafts on the ornithological and ichthyological fauna of the area, it can be proposed to study the reactions of birds and fish without direct measurement of noise as an insitu stimulus. Instead, it is proposed to use "maps" with isolations of noise level for different speeds of hovercrafts prepared in advance according to the measurement data, and distances to the source to be determined by GPS coordinates of the place and GPS tracks of hovercrafts passage. Such estimates cannot replace underwater measurements, but due to their high cost and complexity, such estimates can be useful for preliminary analysis of noise impacts. It should be noted again that the obtained correlation values between airborne and underwater noises refer to the particular hovercraft and acoustic signal propagation conditions.

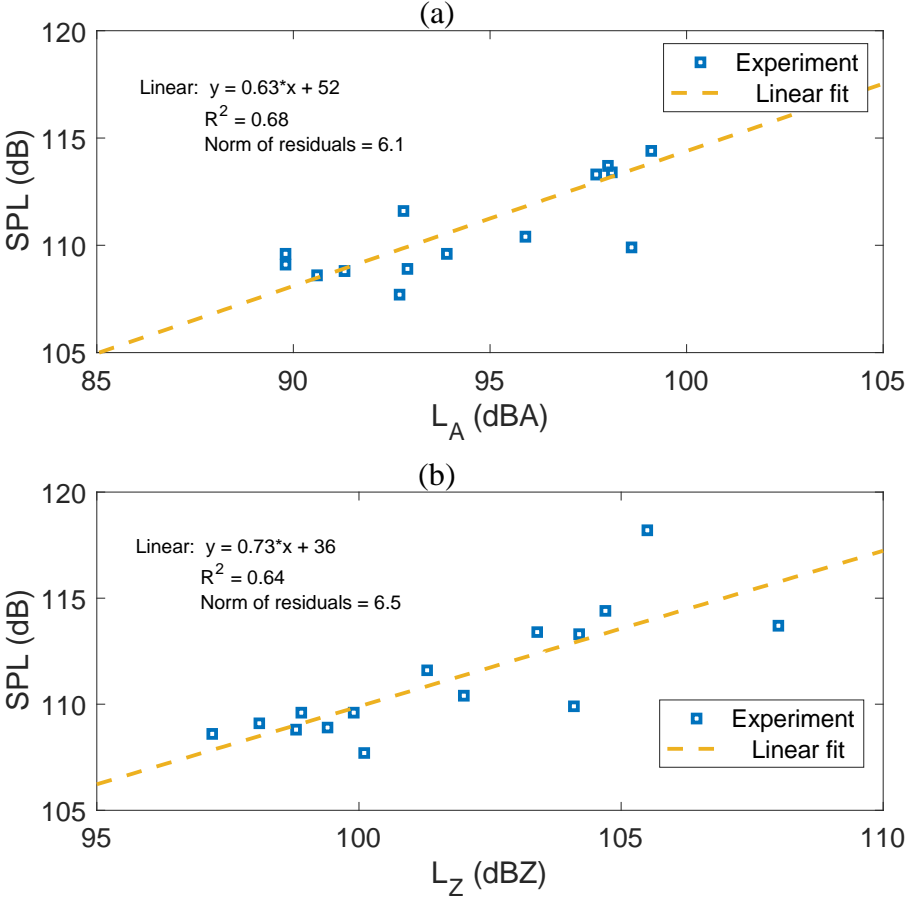

Fig. 8. Maximum level of the underwater noise vs. maximum level of the airborne noise (a) in dBA, (b) in dBZ. One point corresponds to one passage of a hovercraft. Noise levels are normalized to a range of 30 m.

Experimental underwater and airborne noise levels of Griffon BHT130 hovercraft Caspian Falcon demonstrate a strong dependence on its speed, which is determined by the rotation of its engines. The maximum values of noise levels correspond to the hovercraft running at minimum and maximum speeds, and the lowest values for the cruising speed. For any speed, the underwater noise is substantially lower than that radiated by the vessels with screw or water-jet propellers. The airborne noise levels are higher than those of propeller vessels. However, when the hovercraft is moving at the cruising speed (7-15 m/s) in the middle of the Ural-Caspian Channel, the noise level just beyond the shoreline does not exceed the threshold value of 80 dBA for the negative response of birds [*Shannon* *et al., 2016*]. For underwater noise the threshold of 130 dB re 1 µPa has been considered as an onset of the negative behavioral response of fish [*Popper, Technical Report 2014; Hawkins & Popper, 2017*]. It should be noted, that negative noise impact on the wildlife depends on the type of noise (impulsive, transient, continuous), its level and duration of exposure to noise. The majority of studies are carried out to assess the hazardous impact of impulse noise on the aquatic wildlife during seismic surveying or pile driving [*MacGillivray A., Racca R. 2006; R.D. McCauley, M. G. Meekan, M J. G. Parsons, 2021*]. For impulse noise, the thresholds of permissible maximum levels or duration of exposure have been established. Non-impulse (continuous) shipping noise can lead to behavioral effects on fish, such as elicitation of startle response, disruption of feeding, avoidance of an area, masking of the communication signals, disturbance of the migration and spawning. In case of critical levels or prolonged noise exposure, it can cause a temporary or permanent threshold shift of hearing (TTS, PTS) [*Popper A.N., Hawkins A.D., Halvorsen M.B, 2019; Popper, A.N. Hawkins, A.D. et al 2014*]. For the continuous noise impact on behavior, the exposure criteria have not been established yet. For this reason, one of the goal of our work was to measure the parameters of the non-impulsive airborne and underwater noise from the hovercraft in field experiments. The experimentally measured underwater noise levels from hovercraft *Caspian Falcon* did not exceed known threshold. Thus, the potential negative impact of the noise from hovercraft *Caspian Falcon* on the fish fauna within the environmentally sensitive shallow-water Ural-Caspian Channel can be negligible in comparison with the standard ships with underwater engines.

The results obtained in the present work, as well as results of [*Vedenev A.I., Kochetov O.Y. et al. 2023*] refer to the case of hovercraft motion over a free water surface. At the same time, an important question raises about mechanism of the hovercraft noise formation and values of noise levels in the presence of ice cover. In this case the noise generated by hovercraft engines will be added to the noise of ice rushing and cracking, arising when the hovercraft moves over the ice surface. The presence of an additional noise mechanism will increase the overall noise background. While airborne noise levels are likely to retain their value compared to the ice-free case, measurements of acoustic pressure levels generated underwater are of particular interest. The first experiments in this direction are already underway. Further studies of the hovercraft noise fields will be devoted to this issue.

**Acknowledgements**

The authors are grateful to the Personnel of RAS Institute of Oceanology; PGS Designer, Mr. D.A. Shvoyev and Mr. A.V. Shatravin, Member of the experiments held in 2017 and 2019; crew of SD and FRC vessels of Veritas Marine Company; crew members of "Caspian Falcon" vessel of Caspian Offshore Construction Company; personnel of NCOSRB of KMG Systems & Services LLP who ensured implementation of the experiments.

This research was funded by North Caspian Operating Company N.V. (NCOC N.V) grant number JO 229077.